\documentclass[prd,amsmath,amssymb,superscriptaddress,preprintnumbers,twocolumn,10pt]{revtex4-1}

\pdfoutput=1

\usepackage{graphicx}
\usepackage{dcolumn}
\usepackage{bm}
\usepackage{amssymb}
\usepackage{latexsym}
\usepackage{booktabs}
\usepackage{amsmath}
\usepackage{multirow}
\usepackage{url}

\usepackage{float}

\usepackage[normalem]{ulem}
\usepackage{color}
\usepackage{array}
\usepackage{enumerate}
\usepackage[colorlinks=true, linkcolor=red, citecolor=blue]{hyperref}

\newcommand{\aap}{Astron. Astrophys.}

\newcommand{\mnras}{Mon. Not. Roy. Astron. Soc.}

\begin{document}

\title{
Prospects for measuring neutrino mass with 21-cm forest
}

\author{Yue Shao}
\affiliation{Liaoning Key Laboratory of Cosmology and Astrophysics, College of Sciences, Northeastern University, Shenyang 110819, China}

\author{Guo-Hong Du}
\affiliation{Liaoning Key Laboratory of Cosmology and Astrophysics, College of Sciences, Northeastern University, Shenyang 110819, China}

\author{Tian-Nuo Li}
\affiliation{Liaoning Key Laboratory of Cosmology and Astrophysics, College of Sciences, Northeastern University, Shenyang 110819, China}

\author{Xin Zhang}\thanks{Corresponding author}\email{zhangxin@mail.neu.edu.cn}
\affiliation{Liaoning Key Laboratory of Cosmology and Astrophysics, College of Sciences, Northeastern University, Shenyang 110819, China}
\affiliation{National Frontiers Science Center for Industrial Intelligence and Systems Optimization, Northeastern University, Shenyang 110819, China}
\affiliation{MOE Key Laboratory of Data Analytics and Optimization for Smart Industry, Northeastern University, Shenyang 110819, China}

\begin{abstract}
Both particle physics experiments and cosmological observations have been used to explore neutrino properties. Cosmological researches of neutrinos often rely on the early-universe cosmic microwave background observations or other late-universe probes, which mostly focus on large-scale structures. We introduce a distinct probe, the 21-cm forest, that differs from other probes in both time and scale. Actually, the 21-cm forest is a unique tool for studying small-scale structures in the early universe. Below the free-streaming scale, massive neutrinos suppress the matter power spectrum, influencing small-scale fluctuations in the distribution of matter. The one-dimensional (1D) power spectrum of the 21-cm forest can track these fluctuations across different scales, similar to the matter power spectrum, providing an effective method to constrain neutrino mass. Although heating effects in the early universe can also impact the 1D power spectrum of the 21-cm forest, we assess the potential of the 21-cm forest as a tool for measuring neutrino mass, given that the temperature of the intergalactic medium can be constrained using other methods within a certain range. In the ideal scenario, the 21-cm forest observation will have the ability to constrain the total neutrino mass to around 0.1 eV. With the accumulation of observational data and advancements in observational technology, the 21-cm forest holds great promise as an emerging and potent tool for measuring neutrino mass.
\end{abstract}

\maketitle

\section{Introduction}
The phenomenon of neutrino oscillation provides evidence that neutrinos have mass \cite{Maltoni:2004ei}. Indirect measurements of neutrino mass come from solar and atmospheric neutrino experiments, which have measured the squared mass differences of different neutrino masses \cite{Super-Kamiokande:1998kpq,SNO:2002tuh,ParticleDataGroup:2014cgo,DayaBay:2022orm}. In contrast, the $\beta$-decay experiments offer a direct measurement of the neutrino mass. The latest results from the KATRIN collaboration, achieved through precise measurements of the tritium $\beta$-decay spectrum, have established an upper limit for the effective mass, with $m_\beta < 0.45 {\rm~eV}$ \cite{KATRIN:2021uub,Katrin:2024tvg}.

Neutrinos affect the formation and evolution of the large-scale structure of the universe, providing a different method for studying neutrinos compared to particle physics experiments \cite{Lesgourgues:2006nd,Abazajian:2011dt,TopicalConvenersKNAbazajianJECarlstromATLee:2013bxd}. Numerous cosmological observations are employed to constrain the neutrino mass, with the most significant constraints arising from the cosmic microwave background (CMB) \cite{DeBernardis:2009di,WMAP:2010qai,Planck:2018vyg,DiValentino:2024xsv}, which represents the observation of the very early universe. For individual CMB observations, the latest constraint from the Planck satellite data places the upper limit on the total neutrino mass at $\sum m_{\nu} < 0.24 \mathrm{~eV}$ with $95\%$ confidence \cite{Planck:2018vyg}.
Baryon acoustic oscillations (BAO), a late-universe observation from the large-scale structure of the universe, can also be utilized to probe neutrinos. The latest BAO data from the Dark Energy Spectroscopic Instrument, when combined with CMB, has helped to tighten the constraints on the total neutrino mass to $\sum m_{\nu} < 0.072 \mathrm{~eV}$ at the $95\%$ confidence level \cite{DESI:2024mwx}.
The Lyman-$\alpha$ (Ly$\alpha$) forest, another late-universe probe, is important for probing the small-scale structure of the universe and can be used to explore neutrinos \cite{Viel:2005ha,Seljak:2006bg,Gratton:2007tb,Yeche:2017upn,Palanque-Delabrouille:2019iyz,Sarkar:2023pap}. By combining the Ly$\alpha$ forest with the CMB, the latest constraint is $\sum m_{\nu} < 0.11 \mathrm{~eV}$ \cite{Palanque-Delabrouille:2019iyz}.
Moreover, several emerging late-universe cosmological probes, such as gravitational waves \cite{Wang:2018lun,Jin:2022tdf,Feng:2024lzh,Feng:2024mfx} and 21-cm observations \cite{McQuinn:2005hk,Visbal:2008rg,Mao:2008ug,Oyama:2015gma,Zhang:2019ipd}, have also been applied to forecast the capabilities for measuring neutrino mass. Many works have also attempted to constrain the neutrino mass individually or by combining different datasets collectively \cite{Wang:2005vr,Battye:2013xqa,Palanque-Delabrouille:2014jca,Zhang:2014nta,Zhang:2014dxk,Zhang:2014ifa,Zhang:2015rha,Zhang:2015uhk,DellOro:2015kys,Wang:2016tsz,Zhao:2016ecj,Capozzi:2017ipn,Vagnozzi:2017ovm,Guo:2017hea,Zhang:2017rbg,Feng:2017nss,Zhao:2017urm,Feng:2017mfs,Zhao:2017jma,Zhang:2017ljh,Feng:2017usu,Guo:2018gyo,Zhao:2018fjj,Feng:2019mym,Feng:2019jqa,Zhang:2020mox,Li:2020gtk,Feng:2021ipq,Liu:2023qkf,DiValentino:2023fei,Du:2024pai,Toda:2024uff,Reboucas:2024smm,Racco:2024lbu}.

However, while the CMB offers a glimpse into the very early universe, other probes are based on observations in the late universe. For the early universe, particularly during the epoch of reionization, we still lack methods to detect neutrinos. Here we propose a unique probe, known as the 21-cm forest, which is composed of absorption lines at the 21-cm wavelength observed in the spectra of high-redshift radio-bright sources  \cite{Carilli:2002ky, Furlanetto:2002ng, Furlanetto:2006dt, Xu:2009dr, Xu:2010br, Xu:2010us, Ciardi:2012ik}. This probe can investigate the small-scale structures of the early universe, which is different from other probes in terms of both the time, as it focuses on the early universe, and the scale, as it examines smaller scales than other probes, even smaller than those of the Ly$\alpha$ forest.
Furthermore, the one-dimensional (1D) power spectrum of the 21-cm forest reflects fluctuations in the small-scale 21-cm signal within the universe, much like how the matter power spectrum reflects the distribution of matter across various scales \cite{Thyagarajan:2020nch,Shao:2023agv,Sun:2024ywb,Shao:2024owi,Soltinsky:2024mzy}. Since the total neutrino mass suppresses the matter power spectrum on small scales below the free-streaming scale \cite{Hu:1997mj}, this suppression is also reflected in the 1D power spectrum of the 21-cm forest.

In this work, we explore the impact of neutrino mass on the 1D power spectrum of the 21-cm forest and how to use this probe to constrain the total neutrino mass. Note that the heating effect caused by the formation of the first galaxies in the early universe leads to an increase in the temperature of intergalactic medium (IGM) \cite{Xu:2009dr,Mack:2011if,Shimabukuro:2014ava}, which thereby affects the 1D power spectrum of the 21-cm forest. This effect is degenerate with the influence of neutrino mass, making it difficult to distinguish between the two. Therefore, we further assess the ability of the 1D power spectrum of the 21-cm forest to constrain the total neutrino mass, given that other methods can precisely constrain the IGM temperature.
We also discuss the potential of the 21-cm forest to constrain the total neutrino mass compared with other probes, if we are able to observe more neutral segments of the 21-cm forest in the future.

The paper is organized as follows. In Section~\ref{sec_nm}, we introduce the effect of the total neutrino mass on the suppression of the matter power spectrum. In Section~\ref{sec_hmf}, we analyze the impact of total neutrino mass on the halo mass function. In Section~\ref{sec_1dps}, we provide an analytical model for the 1D power spectrum of the 21-cm forest. In Section~\ref{sec_noise}, we evaluate the observational uncertainty of the 1D power spectrum. In Section~\ref{res}, we present the results of measuring the total neutrino mass using the 21-cm forest, followed by the conclusion in Section~\ref{con}. The cosmological parameters relevant to this work are adopted from Planck 2018 \cite{Planck:2018vyg}.

\section{Method}\label{method}
\subsection{Neutrino mass}\label{sec_nm}
The presence or absence of mass in neutrinos significantly affects their cosmological behavior. In the early universe, massless neutrinos behave primarily as a radiation component, with their energy density evolving in accordance with the radiation. In contrast, after decoupling, massive neutrinos act more like matter, with their energy density evolving similarly to that of ordinary matter. Due to their non-zero mass, these neutrinos suppress structure formation below a characteristic free-streaming scale \cite{Hu:1997mj}.
Below this scale, neutrinos can affect the clustering of matter, reducing the amplitude of density perturbations. This results in a suppression of the matter power spectrum on small scales, and the degree of suppression can be quantitatively described by \cite{Hu:1997mj}
\begin{equation}\label{dp_p}
\left(\frac{\Delta P}{P}\right) \approx -8 \frac{\Omega_\nu}{\Omega_{\rm m}} \approx -\left(\frac{0.08}{\Omega_{\rm m} h^2}\right) \left(\frac{\sum m_{\nu}}{1 \mathrm{~eV}}\right),
\end{equation}
where $\sum m_{\nu}$ denotes the total neutrino mass, $\Omega_{\rm m}$ is baryon density parameter, and $h$ is dimensionless Hubble constant. $\Omega_\nu$ represents the ratio of the energy density of massive neutrinos $\rho_\nu$ to the critical density $\rho_{\rm cr}$, and is calculated as
\begin{equation}
\Omega_\nu=\frac{\rho_\nu}{\rho_{\rm cr}}=\frac{\sum m_{\nu}}{93.14~h^2\mathrm{~eV}}.
\end{equation}
By influencing the matter power spectrum, massive neutrinos also reduce the formation rate of low-mass dark matter halos, thus affecting the halo mass function \cite{Shimabukuro:2014ava}.

\subsection{Halo mass function}\label{sec_hmf}
We use the Press-Schechter (PS) halo mass function \cite{Press:1973iz,Barkana:2000fd} to describe the number of dark matter halos within a given mass range per unit volume,
\begin{align}
\frac{{\rm d} n(z, M)}{{\rm d} M} = \sqrt{\frac{2}{\pi}} \frac{\rho_{\rm m}}{M} \left|\frac{{\rm d} \sigma(M)}{{\rm d} M}\right| \frac{\delta_{\rm c}(z)}{\sigma^3(M)} \exp \left[ -\frac{\delta^2_{\rm c}(z)}{2\sigma^2(M)} \right],
\end{align}
where $\rho_{\rm m}$ is the current average matter density in the universe, and $\delta_{\rm c}(z)$ is the critical overdensity for collapse. $\sigma(M)$ represents the standard deviation of the mass distribution \cite{Barkana:2000fd}, given by
\begin{equation}
\sigma^2(M) = \sigma^2(R) = \frac{1}{2 \pi^2} \int k^2 P(k)\left[\frac{3 j_1(k R)}{k R}\right]^2 {\rm d} k,
\end{equation}
where $M = 4\pi \rho_{\rm m} R^3 /3$ with $R$ being the comoving radius, $j_1(x) = (\sin x - x \cos x) / x^2$, and $P(k)$ is matter power spectrum mentioned in Eq.~(\ref{dp_p}).

\begin{figure}
\centering
\includegraphics[angle=0, width=8cm]{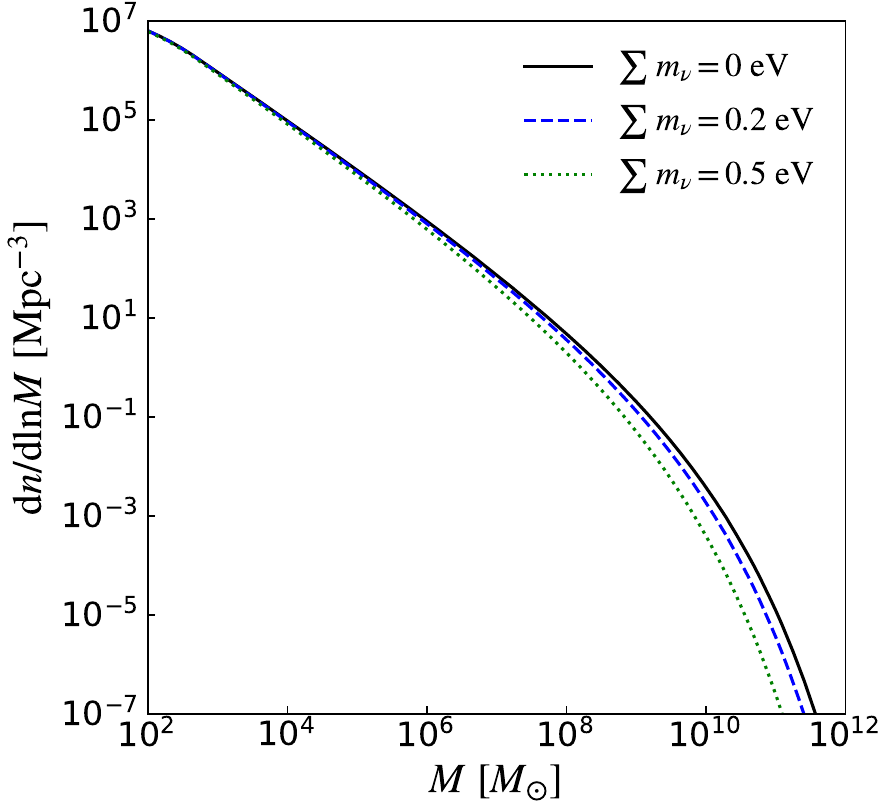}
\caption{\label{hmf}
  Halo mass function for different values of the total neutrino mass at $z = 9$. The black, blue and green curves correspond to $\sum m_{\nu} = 0 {\rm ~eV}$, $\sum m_{\nu} = 0.2 {\rm ~eV}$ and $\sum m_{\nu} = 0.5 {\rm ~eV}$, respectively.
}
\end{figure}

We show the halo mass function in Fig.~\ref{hmf}. We can see that the greater the total neutrino mass, the more pronounced the suppression on the halo mass function.

\subsection{1D power spectrum of 21-cm forest}\label{sec_1dps}
We employ the halo model approach \cite{Ma:2000ik,Smith:2002dz,Mead:2015yca,Philcox:2020rpe,Acuto:2021yjm,Feng:2017ttq, Schneider:2020xmf,Schneider:2023ciq,Schaeffer:2023rsy,Hitz:2024cwl} to directly model the 1D power spectrum of the 21-cm forest \cite{Shao:2024owi}, thereby circumventing the complex simulations involved in the optical depth and brightness temperature of the 21-cm forest. The 1D power spectrum of the 21-cm forest brightness temperature can be expressed as
\begin{align}\label{p21}
P(k_{\|}, z) = T_0^2(z) P_{21}(k_{\|}, z),
\end{align}
where $P_{21}(k_{\|}, z)$ and $T_0(z)$ represent the scale-dependent part and scale-independent part, respectively. $T_0(z) = - 0.0085 (1+z)^{1/2} T_{\gamma}(z)$ is only related to the brightness temperature of the background radiation, consisting of the background point source temperature $T_{\mathrm{point}}$ and the CMB temperature.
The observed brightness temperature of a background source is related to the flux density $S_{150}$ by
\begin{align}
T_{\mathrm{point}}(\nu, z=0) = \frac{S_{150} c^2}{2 k_{\rm B} \nu^2 \Omega} \left( \frac {\nu}{\nu_{150}} \right)^\eta,
\end{align}
where $S_{150}$ represents the flux density at $150 {\rm~MHz}$, $c$ is the speed of light, $k_{\rm B}$ is the Boltzmann constant, and $\eta = -1.05$ \cite{Carilli:2002ky}. The solid angle $\Omega$ is given by $\Omega = \pi(\theta/2)^2$, where $\theta = 1.22 \lambda_z/D$ is the angular resolution, with $\lambda_z$ being the observed wavelength and $D$ the longest baseline of the telescope. We take $D = 65 {\rm~km}$ for the low-frequency array of the Square Kilometre Array (SKA-LOW).

The scale-dependent part $P_{21}(k_{\|}, z)$ is the 1D power spectrum of $[1+\delta(z)]/T_{\rm S}(z)$, where $\delta$ and $T_{\rm S}$ represent the gas overdensity and the spin temperature, respectively. By assuming that spin temperature is coupled to gas kinetic temperature $T_{\rm K}$ by the early Ly$\alpha$ background, it can be written as $[1+\delta(z)]/T_{\rm K}(z)$. These are the primary perturbations on small scales in the early universe. We project the average three-dimensional (3D) 21-cm power spectrum along the line of sight to calculate the 1D power spectrum,
\begin{align}\label{pave}
P_{21}(k_{\|}, z) = P_{21,{\rm 1D}}(k, z) = \frac{1}{2 \pi} \int_k^{\infty} k' P_{21}(k', z) {\rm d}k'.
\end{align}
The averaged 3D 21-cm power spectrum $P_{21}(k, z)$ is decomposed into $P_{21}^{1h}(k, z)$ and $P_{21}^{2h}(k, z)$, which account for the influence of individual halos and the interaction between pairs of halos. These components can be written as
\begin{equation}
\begin{aligned}
P_{21}^{1h}(k, z) &= \frac{1}{\langle \rho_{21} \rangle^2} \int {\rm d} M \frac{{\rm d} n}{{\rm d} M} |W_{21}|^2, \\
P_{21}^{2h}(k, z) &= \frac{1}{\langle \rho_{21} \rangle^2}
\left[ \int {\rm d} M \frac{{\rm d} n}{{\rm d} M} |W_{21}| b \right]^2 \times P_{\rm lin}.
\end{aligned}
\end{equation}
Here, ${\rm d} n(z, M) / {\rm d} M$ is the halo mass function introduced in the previous section, $b(z, M)$ is the halo bias \cite{Mo:1995cs}, and $P_{\rm lin} (k, z)$ represents the linear matter power spectrum. The window function $W_{21}(k, z, M)$ and the average density profile $\langle \rho_{21}(z) \rangle$ are related to the profile $\rho_{21}(r, z, M)$,
\begin{equation}
\begin{aligned}
&W_{21}(k, z, M) = 4 \pi \int {\rm d} r r^2 \rho_{21}(r, z, M) \frac{\sin(k r)}{k r}, \\
&\langle \rho_{21}(z) \rangle = 4 \pi \int {\rm d} M \frac{{\rm d} n(z, M)}{{\rm d} M} \int {\rm d} r r^2 \rho_{21}(r, z, M),
\end{aligned}
\end{equation}
where $\rho_{21}(r, z, M)$ is the profile of  $[1+\delta(z)]/T_{\rm K}(z)$, consisting of the density profile and the temperature profile.

We use Navarro-Frenk-White \cite{Navarro:1996gj} profile to describe the profile of dark matter, and the gas density profile can be analytically derived as follows \cite{Makino:1997dv}
\begin{align}
\ln \rho_{\rm g}(r) = \ln \rho_{\rm gc} - \frac{\mu m_{\rm p}}{2 k_{\rm B} T_{\rm vir}} \left[ v_{\rm e}^{2}(0) - v_{\rm e}^{2}(r) \right],
\end{align}
where $\rho_{\rm g}$ and $\rho_{\rm gc}$ are gas density at radius $r$ and at the center, $v_{\rm e}$ is the gas escape velocity, $\mu$ is the mean molecular weight, $m_{\rm p}$ is the proton mass, and $T_{\rm vir}$ is the virial temperature.
Outside the virial radius, we use the infall model \cite{Barkana:2002bm} to describe the gas inflow process towards the halo caused by gravitational effects. The infall model also describes the distribution of dark matter, and here we assume that the gas density profile is consistent with the dark matter distribution.

For the temperature profile, within the virial radius, we consider the gas temperature to be equal to the virial temperature of the halo. Outside the virial radius, in the IGM, the gas temperature is determined by the balance between adiabatic cooling due to cosmic expansion and heating from X-rays. When X-ray heating is strong, the IGM temperature is dominated by X-ray heating. We consider an IGM temperature of $T_{\rm K} = 60 {\rm~K}$, which is a reasonable value within the observational constraints of Hydrogen Epoch of Reionization Array \cite{HERA:2021noe,HERA:2022wmy}.

\subsection{Observational uncertainty}\label{sec_noise}
The thermal noise on the 1D power spectrum is estimated by \cite{Thyagarajan:2020nch,Shao:2023agv},
\begin{align}
P^N = \frac{1}{\sqrt{N_s}} \left(\frac{\lambda_z^2 T_{\mathrm{sys}}}{ A_{\rm {eff }} \Omega}\right)^2\left(\frac{\Delta r_z}{ \Delta \nu_z \delta t}\right),
\end{align}
where $N_s$ is the number of neutral segments, $\Delta \nu_z$ is the total observing bandwidth corresponding to $\Delta r_z$, $\delta t$ represents the integration time, and $A_{\rm eff }/T_{\rm sys}$ is the sensitivity of telescope. For 21-cm forest observations at suitable redshifts, SKA-LOW can achieve a sensitivity of $500-600{\rm~m^2~K^{-1}}$ \cite{Braun:2019gdo}. In addition, the sample variance should also be taken into account within the observational uncertainty, and we use $P (k_{\|}, z)$ to estimate the sample variance.

\section{Results and discussion}\label{res}

\begin{figure}
\centering
\includegraphics[angle=0, width=8cm]{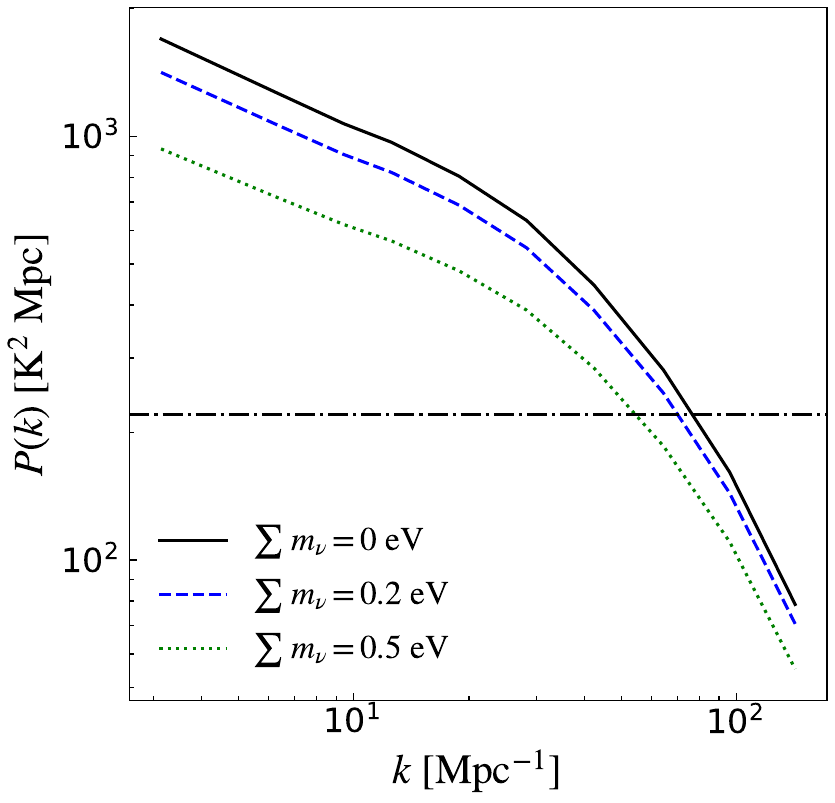}
\caption{\label{ps}
  The 1D power spectrum of the 21-cm forest with different total neutrino mass at $z = 9$. The black, blue and green curves correspond to $\sum m_{\nu} = 0 {\rm ~eV}$, $\sum m_{\nu} = 0.2 {\rm ~eV}$ and $\sum m_{\nu} = 0.5 {\rm ~eV}$, respectively. The black dash-dot line corresponds to the thermal noise of the SKA-LOW observed over 100 h, assuming an average over 100 segments.
}
\end{figure}

In this work, we consider the 1D power spectrum corresponding to a 21-cm forest signal along the line of sight with a length of 2 Mpc. We focus on the behavior of the 21-cm forest on small scales, as the analytical model we used has been shown to match well with the simulation results when only small-scale neutral regions are considered \cite{Shao:2024owi}. For modeling on larger scales, it is necessary to incorporate the ionization profile. However, there is no 21-cm forest signal within ionized regions, and its impact on the 1D power spectrum remains unclear. We adopt a frequency resolution of 1 kHz, which corresponds to $k = 356 {\rm ~Mpc}^{-1}$ at redshift 9 and represents the smallest scale that can be probed by the 1D power spectrum.

In Fig.~\ref{ps}, we show the 1D power spectrum of the 21-cm forest that we model for different total neutrino masses, assuming $T_{\rm K} = 60 {\rm ~K}$.
We can see that, similar to the suppression of the matter power spectrum by the total neutrino mass, the 1D power spectrum of the 21-cm forest is also suppressed. Therefore, the 1D power spectrum can serve as a new observational tool to constrain the total neutrino mass.

The heating effect in the early universe will also suppress the 1D power spectrum overall, and this effect is highly degenerate with the total neutrino mass impact, making them difficult to distinguish. To estimate the error in the total neutrino mass, we use the Fisher matrix formalism. If the relative error of $T_{\rm K}$ can be determined through other observational methods, a prior on temperature relative error can be introduced into the Fisher matrix to constrain the total neutrino mass.

We take the high-redshift radio-loud quasars as the background sources in the 21-cm forest observations. For a radio-loud quasar, we assume  a flux density of 10 mJy at redshift 9, observing for 100 h. Since we consider neutral segments only 2 Mpc in length, we can extract 100, 1000, or even 10000 segments from the quasar spectrum. Although extracting 1000 or 10000 segments may be challenging, this becomes feasible if more radio-loud quasars can be observed. With the advancement in models for the abundance of high-redshift radio-loud quasars \cite{Haiman:2004ny,Niu:2024eyf} and the discovery of radio-loud quasars at higher redshifts \cite{McGreer:2006ey,Willott:2009wv,Banados:2015fda,Banados:2021imw,Banados:2022mrl,Banados:2024xds,2021A&A...647L..11I,2021ApJ...908..124L,2022A&A...668A..27G,2023MNRAS.519.2060I,2023MNRAS.520.4609E}, this is also something to look forward to.

\begin{figure*}
\centering
\includegraphics[angle=0, width=16cm]{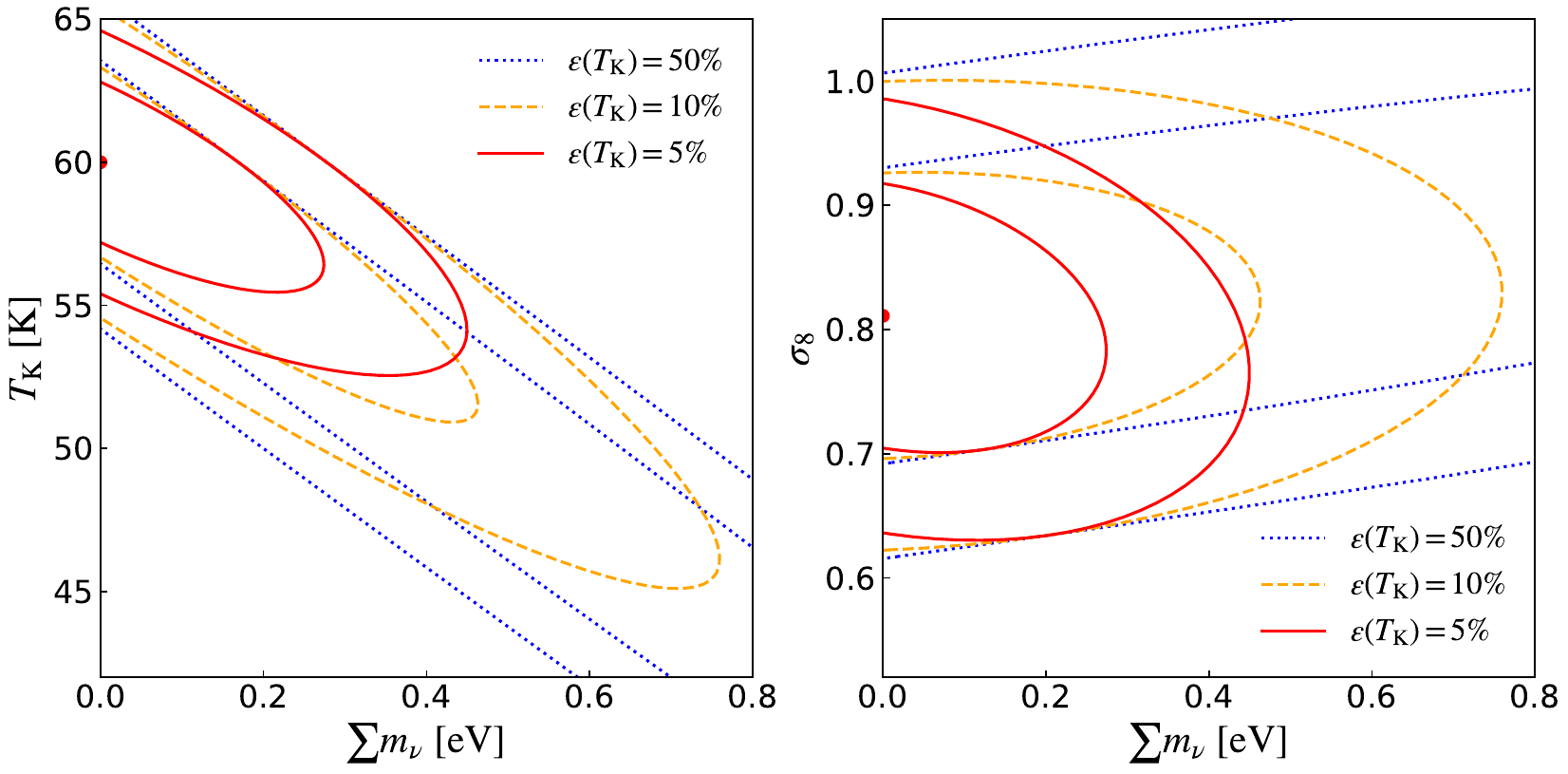}
\caption{\label{contour}
 Prospective constraints on the neutrino mass with the 21-cm forest. The left panel shows the constrains on $\sum m_{\nu}$ and $T_{\rm K}$ with confidence levels of $68.3\%$ and $95.4\%$. The right panel shows the constrains on $\sum m_{\nu}$ and $\sigma_8$ with same confidence levels. The red, yellow and blue contours correspond to the prior of $\varepsilon (T_{\rm K})=5\%$, $\varepsilon (T_{\rm K})=10\%$ and $\varepsilon (T_{\rm K})=50\%$, respectively.}
\end{figure*}

We present the predicted contours based on 100 21-cm forest segments in Fig.~\ref{contour}. In the left panel, the blue ellipse represents a prior of $\varepsilon (T_{\rm K})=50\%$, where a degeneracy exists between the effects of the IGM temperature and the total neutrino mass on the 1D power spectrum, which makes it challenging to distinguish between them. If other probes can reduce the relative temperature errors $\varepsilon (T_{\rm K})$ to $10\%$ or even $5\%$, it will help break this degeneracy, leading to more accurate measurements of the total neutrino mass.
The right panel illustrates a degeneracy between total neutrino mass and $\sigma_8$, as both parameters similarly affect the amplitude of the matter power spectrum. Despite insufficient precision, the 21-cm forest can still be used to constrain $\sigma_8$, and holds promise for future improvements to improve the precision of CMB constraints on $\sigma_8$.

Note that in addition to the degeneracy with total neutrino mass for $T_{\rm K}$ and $\sigma_8$, we also consider $H_0$ in the Fisher matrix and use the prior from Planck 2018 ($H_0 = 67.36 \pm 0.6$ km s$^{-1}$ Mpc$^{-1}$). Although the 21-cm forest may not be sensitive to $H_0$, we hope that future 21-cm forest observations can provide a $H_0$ value from the early-universe observation. This would help alleviate the $H_0$ tension between very early-universe CMB observations and various late-universe cosmological probes (for the Hubble tension, see, e.g., Refs.~\cite{Li:2013dha,Zhang:2014dxk,Bernal:2016gxb,Zhao:2017urm,Guo:2018ans,Verde:2019ivm,Riess:2019qba,Vagnozzi:2019ezj,DiValentino:2021izs,Gao:2021xnk,Gao:2022ahg,Vagnozzi:2023nrq,Huang:2024erq}). In addition, we fix other cosmological parameters.
Moreover, in the calculations of 1D power spectrum of 21-cm forest and the linear matter power spectrum, as well as in the analysis of the Fisher matrix, we fix $\Omega_{\rm m} = 0.3153$.

\begin{figure*}
\centering
\includegraphics[angle=0, width=16cm]{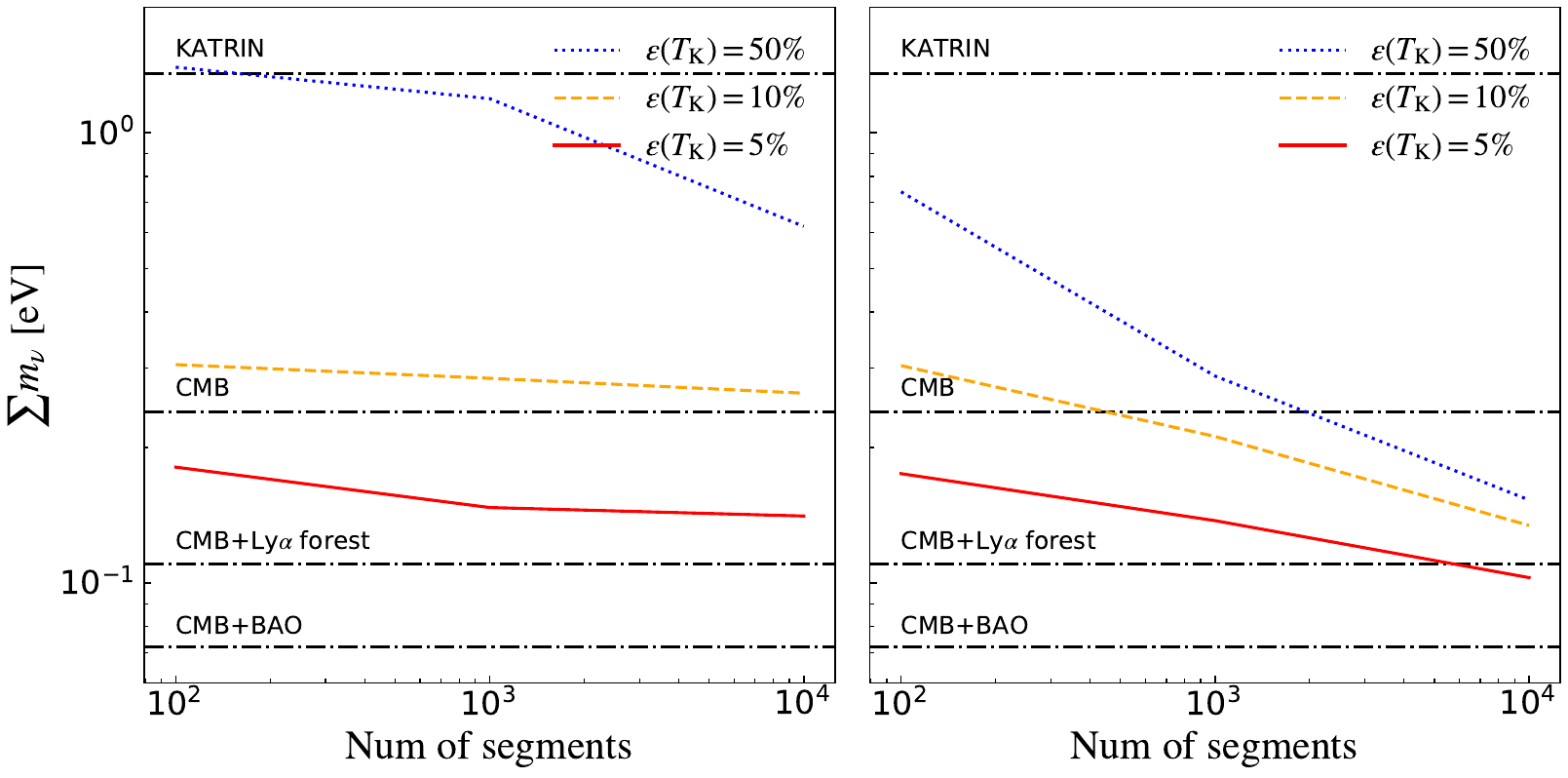}
\caption{\label{err}
  Upper limit on the total neutrino mass for different numbers of segments. The left panel shows the results without considering the prior on $\sigma_8$, whereas the right panel incorporates the prior on $\sigma_8$ from Planck 2018. The red, yellow and blue lines correspond to priors on the relative temperature error, with $\varepsilon (T_{\rm K})=5\%$, $\varepsilon (T_{\rm K})=10\%$, and $\varepsilon (T_{\rm K})=50\%$, respectively. The black line represents the constraints from KATRIN, CMB, CMB+Ly$\alpha$ forest, and CMB+BAO, respectively.
}
\end{figure*}

In Fig.~\ref{err}, we show the upper limits on the total neutrino mass, assuming different numbers of observable neutral segments in the 21-cm forest and varying relative error of $T_{\rm K}$.
The left panel shows that if the relative error of $T_{\rm K}$ can reach $5\%$, only 100 segments can constrain the total neutrino mass to $0.18 {\rm ~eV}$, which is better than CMB. If the relative error of $T_{\rm K}$ is $10\%$, the 21-cm forest still has the ability to constrain the neutrino mass close to the level of the CMB.
However, as the number of 21-cm forest segments increases, the ability of the 21-cm forest to constrain the total neutrino mass does not improve significantly, due to the degeneracy between $\sigma_8$ and $\sum m_{\nu}$. Therefore, in the right panel, we incorporate a prior from Planck 2018 ($\sigma_8 = 0.8111 \pm 0.0073$) to estimate the capability of future 21-cm forest observations in combination with CMB. If the relative error of $T_{\rm K}$ is $50\%$, the result of 1000 segments is $\sum m_{\nu} < 0.29 {\rm~eV}$. As the relative error decreases and the number of segments increases, the constraint of the 21-cm forest on the total neutrino mass will approach $0.1 {\rm ~eV}$,  comparable to the Ly$\alpha$ forest.
Furthermore, $0.1 {\rm ~eV}$ is also the lower limit for the inverted hierarchy of neutrino masses, while the lower limit for the normal hierarchy is $0.06 {\rm ~eV}$. If future 21-cm forest observations in combination with CMB or other cosmological probes can constrain the total neutrino mass to $0.1 {\rm ~eV}$, it would further enable an effective distinction between neutrino mass hierarchies.

We wish to further clarify the origin of the inflection at $N_s=1000$ in Fig.~\ref{err}. As demonstrated in Fig.~\ref{ps}, our analysis of 100 segments reveals thermal noise dominance at scales around $k=60 \mathrm{~Mpc}^{-1}$ for 21-cm forest observations. When increasing the segment number to 1000, the thermal noise decreases, extending our detectable range to smaller spatial scales (higher $k$-modes) in the 1D power spectrum. This improvement in sensitivity enables the detection of finer-scale structures, but introduces two competing effects: while reduced noise enhances measurement precision, the intrinsic signal amplitude diminishes significantly at these smaller scales. Furthermore, instrumental limitations in frequency resolution impose additional constraints on resolving ultra-fine features. The signal-to-noise ratio (SNR) evolution thus represents a balance between these factors. Beyond $N_s=1000$, although continued noise reduction theoretically permits access to progressively smaller scales, the combination of diminishing 1D power spectrum amplitude and resolution limitations causes the SNR enhancement to plateau. However, we emphasize that the precise position of this inflection point remains observationally contingent, as current uncertainties in critical astrophysical parameters --- particularly the IGM temperature and background source flux densities --- introduce substantial variance in the signal amplitude predictions.

At last, we make a brief discussion on the choice of halo mass function. We employed the classical PS halo mass function in our calculations. This choice is particularly appropriate for analyzing the primordial dark matter halos probed by the 21-cm forest signal, as these low-mass structures predominantly exist in a nearly spherical state in the early universe and do not form galaxies --- under such conditions, the PS framework will maintain its theoretical validity. To quantify potential systematic uncertainties, we conducted comparative analyses with the Sheth-Tormen (ST) mass function \cite{Sheth:2001dp}, which incorporates ellipsoidal collapse dynamics. Our numerical tests with 1000 spectral segments reveal remarkable consistency: under a $50\%~ T_{\mathrm{K}}$ uncertainty, the derived neutrino mass upper limits show only $10\%$ variation ( $\sum m_{\nu}<0.32 \mathrm{~eV}$ vs. $0.29 \mathrm{~eV}$ for ST and PS respectively). This minimal discrepancy demonstrates the relative insensitivity of our constraints to halo collapse model selection, reinforcing the robustness of our PS-based approach for this specific cosmological regime.

\section{Conclusion}\label{con}
The 21-cm forest is one of the few probes capable of detecting small-scale structures in the early universe. The 1D power spectrum reflects fluctuations in the 21-cm signal on small scales. Massive neutrinos suppress the formation of structures, making it possible to detect their effects using the 1D power spectrum of the 21-cm forest. Using radio-bright background sources, such as radio-loud quasars, the 21-cm forest, with its high SNR, offers a promising tool for measuring total neutrino mass.

If future measurements can precisely constrain the temperature of the IGM, the 21-cm forest has the potential to constrain the total neutrino mass to $0.18 {\rm ~eV}$.
Despite the increase in neutral segments, the constraints on the total neutrino mass do not show a substantial improvement due to the degeneracy between $\sigma_8$ and $\sum m_{\nu}$. By incorporating a prior on $\sigma_8$, we can enhance the limitations on total neutrino mass, possibly reaching around $0.1 {\rm ~eV}$. This enhancement will offer vital clues for investigating the neutrino mass hierarchy.

In this work, we propose a unique method for measuring the total neutrino mass using 21-cm forest observations. This method can explore the impact of neutrinos on the small-scale structure of the early universe and complements other probes in both time and scale. The 21-cm forest leverages the impact of the total neutrino mass on the small-scale fluctuations in the universe, which suppresses the matter power spectrum. This suppression is clearly reflected in the 1D power spectrum of the 21-cm forest.
However, our estimates are still preliminary. With the implementation of 21-cm forest observations and the combination with other early or late universe probes, 21-cm forest observations are expected to offer new insights into neutrino physics.

\section*{Acknowledgments}
This work was supported by the National SKA Program of China (grant Nos. 2022SKA0110200 and 2022SKA0110203), the National Natural Science Foundation of China (grant Nos. 12473001, 11975072, 11875102, and 11835009), and the National 111 Project (Grant No. B16009).

\bibliography{main}

\end{document}